\documentclass[12pt]{iopart}
\usepackage{iopams}
\usepackage{setstack}
\usepackage{graphicx}

\begin{document}

\title{Persistent Homology of Complex Networks}

\author{Danijela Horak$^{1,2}$, Slobodan Maleti\'{c}$^{1}$and Milan Rajkovi%
\'{c}$^{1,\ast }$}

\address{$^{1}$Institute of Nuclear Sciences Vin\v{c}a, Belgrade 11001,
Serbia,\\
$^{2}$Max Planck Institute for Mathematics in the Natural Sciences, \\
\ 04103, Leipzig, Germany}

\ead{*milanr@vin.bg.ac.yu}
\begin{abstract}
Long lived topological features are distinguished from short lived
ones (considered as topological noise) in simplicial complexes
constructed from complex networks. A new topological \ invariant,
persistent homology, is determined and presented as a parametrized
version of a Betti number. Complex networks with distinct degree
distributions exhibit distinct persistent topological features.
Persistent topological attributes, shown to be related to robust
quality of networks, also reflect defficiency in certain
connectivity properites of networks. Random networks, networks with
exponential conectivity distribution and scale-free networks were
considered for homological persistency analysis.

\end{abstract}

\maketitle

\section{Introduction}
Complex systems consisting of large number of highly interconnected
dynamic units, whose structure is usually irregular have been the
subject of intense research efforts in the past few years
\cite{stefano}. The complexity of such systems is reflected not only
in their structure but also in their dynamics. The usual
representation of a wide range of systems of this kind in nature and
society uses networks as the concept appropriate for the study of
both the topology and dynamics of complex systems. The usual
approach to study networks is via graph theory which was well
developed for regular and random graphs both of which have been
found to be exceptional cases of limited use in real world
realizations and applications. Recently, along with the discovery of
new types of network structures such as the small-world \cite{watts}
and scale-free networks \cite{barabasi1}, the tools of statistical
mechanics have been successfully implemented offering explanations
and insights into the the newly recognized properties of these
systems. In spite of many advances based on statistical mechanics
approaches to various issues involving networks, from biology to
social sciences, it is our opinion that there is a need for more
versatile approach which would rely on new topological methods
either separately or in combination with the techniques of
statistical mechanics. In particular, the program is to encode the
network into a simplicial complex which may be considered as a
combinatorial version of a topological space whose properties may
now be studied from combinatorial, topological or algebraic aspects.
The motivation stems from the Q-analysis introduced by R. Atkin
\cite{atkin1}, \cite{atkin2} who advocated its use in various areas
of physics and social systems analysis in the 70's. The methods of
Q-analysis were extended further into a combinatorial homotopy
theory, called A-theory \cite{barcelo}. Consequently, the invariants
of simplicial complexes may be defined from three different points
of view (combinatorial, topological or algebraic) and each one of
them provides completely different measures of the complex and, by
extension, of the graph (network) from which the complex was
constructed. In \cite{milanr}, for several standard types of
networks we constructed vector valued quantities representing
topological and algebraic invariants and showed, among other issues,
that their statistical properties perfectly match their
corresponding degree distributions. Such an approach provided a link
between topological properties of simplicial complexes and
statistical mechanics of networks from which simplicial complexes
were constructed.

In the present exposition we focus on simplicial complexes (obtained
from random, scale- free networks and networks with exponential
conectivity distributions) and their homological properties. In most
general terms, algebraic topology offers two methods for gauging the
global properties of a particular topological space $X$ by
associating with it a collection of algebraic objects. The first set
of invariants are the \textit{homotopy groups} $\pi _{i}(X),$
$i=1,2,...,$ the first one (i.e. for $i=1$), known as the
fundamental homotopy group being well known. Homotopy groups contain
information on the number and kind of ways one can map a
$k$-dimensional sphere $S^{k}$ into $X$, with two spheres in $X$
considered equivalent if they are homotopic (belonging to a same
path equivalence class) relative to some fixed basepoint.
Computational demands of such an approach are in general extremely
high and for that reason the second set of invariants, the
\textit{homology groups, }is of more practical interest. Homology
groups of dimension $k$, $H_{k}(X)$, provide information about
properties of chains formed from simple oriented units known as
simplices. The elements of
homology groups are cycles (chains with vanishing boundary) and two $k$%
-cycles are considered homologous if their difference is the boundary of $%
(k+1)$-chain. In more general terms $H_{k}(X)$ determines the number of $k$%
-dimensional subspaces of $X$ which have no boundary in $X$ and
themselves are not boundary of any $k+1$-dimensional subspace. In
contrast to homotopy groups, homology groups can be computed using
the methods of linear algebra and the ease of these methods are
counterbalanced by obtained topological resolution. It should be
remarked that these computations can be quite time
consuming in spite of recent advances in computational techniques \cite%
{mishaikow}. Although homology groups are computable and provide
insight into topological spaces and maps between them, our interest
is in discerning which topological features are essential and which
can be safely ignored, similar to signal processing procedure when
signal is removed from noise. One of the important informations
about the topological space is the number and type of holes it
contains and going beyond standard homological approaches one could
be interested in finding out which holes are essential and which are
unimportant. This is the subject of persistence and persistent
homology, as introduced by Edelsbrunner, Letscher and Zomorodian \cite{edels}%
, whose aim is to extract long-lived topological features
(topological signal) which persist over a certain parameter range
and which are contrasted with short-lived features (topological
noise).

With networks encoded into simplicial complexes we are interested in
topological features which persist over a sequence of simplicial
complexes of different sizes. This sequence reflects the formation
of the network or the change of the existing network when new node
or nodes are introduced or removed. Here we focus on recognizing
persistent and non persistent features of random, modular and non
modular scale-free networks and networks with exponential
connectivity distribution. In the following exposition our main
topic will be homology and although it is self contained an
elementary knowledge of homology would be helpfull, as may be found
for example in Chapter 2 of \cite{hatcher}. Our main motivation is
to show that each of these different types of networks have
different persistent homological properties although here we do not
attempt to present these features as generic. Moreover, long-lived
topological attributes reveal new and important information related
to connectivity of the network which could not be inferred using any
other conventional methods.

The outline of the exposition is as follows: In Section 2 we review
concepts from algebra and simplicial homology while in Section 3 we
present the methods of constructing simplicial complexes from
graphs. In Section 4 we introduce the concept of persistent homology
and discuss computational aspects. Section 5 contains description of
graphical representation of persistent homology groups. In Section 6
we present the results of persistent homology calculations for
random networks while in Section 7 and 8 persistent homologies are
determined for networks with exponential degree distribution and
three types of scale-free networks respectively. Concluding remarks
are given in Section 9.

\section{Algebraic Topology}

\subsection{Simplicial complexes}

Any subset of $V=$ $\{v_{\alpha _{0}},v_{\alpha _{1}},...,v_{\alpha
_{n}}\}$ determines an $\ n$-$simplex$ denoted by $\left\langle
v_{\alpha
_{0}},v_{\alpha _{1}},...,v_{\alpha _{n}}\right\rangle .$ The elements $%
v_{\alpha _{i}}$ of $V$ are the vertices of the simplex denoted by $%
\left\langle v_{\alpha _{i}}\right\rangle ,$ and $n$ is the
dimension of the simplex. Any set of simplices with vertices in $V$
is called a simplicial
family and its dimension is the largest dimension of its simplices. A $q$%
-simplex $\sigma _{q}$ is a $q$-face of an $n$-simplex $\sigma
_{n}$, denoted by $\sigma _{q}\lesssim \sigma _{n}$, if every vertex
of $\sigma _{q} $ is also a vertex of $\sigma _{n}.$ A simplicial
complex represents a collection of simplices. More formally, a
simplicial complex $K$ on a finite set $V=\{v_{1},...,v_{n}\}$ of
vertices is a nonempty subset of the power set of $V$, so that the
simplicial complex $K$ is closed under the formation
of subsets. Hence, if $\sigma \in K$ and $\rho $.$\in $ $\sigma ,$ then $%
\rho $.$\in K.$

Two simplices $\sigma $ and $\rho $ are $q-connected$ if there is a
sequence of simplices $\sigma ,\sigma _{1},\sigma _{2},...,\sigma
_{n},\rho ,$ such that any two consecutive ones share a
$q$-face,.implying that they have at least $q+1$ vertices in common.
Such a chain is called a $q$-chain. The complex $K$ is $q$-connected
if any two simplices in $K$ of dimensionality
greater or equal to $q$ are $q$-connected. The dimension of a simplex $%
\sigma $ is equal to the number of vertices defining it minus one.
The dimension of the simplicial complex $K$ is the maximum of the
dimensions of the simplices comprising $K$. In Fig. 1 we show an
example of a simplicial
complex and its matrix representation. In this example $V=\{1,2,...,11,12$%
\}, and the simplicial complex $K$ consists of the subsets $%
\{1,2,3,4\},\{3,4,5\},\{5,8\},\{3,6,7\},\{7,8,9,10,11\}$ and $\{9,10,11,12\}$%
. Its dimension is $4$, as there is a $4$-dimensional simplex, in
addition to two $3$-dimensional ones, two $2$-dimensional and one
$1$-dimensional simplex. A convenient way to represent a simplicial
complex is via a so called incidence matrix, whose columns are
labeled by its vertices and whose rows are labeled by its simplices,
as shown also in Fig. 1. The multifaceted property (algebraic,
topological and combinatorial) of simplicial complexes makes them
particularly convenient for modelling complex structures and
connectedness between different substructures.

\subsection{Chains, Cycles and Boundaries}

Chains and cycles are simplicial analogs of paths and loops in the
continuous domain. The set of all $k$-chains together with the
operation of addition forms a group $C_{k}$. A collection of
$(k-1)$-dimensional faces of a $k$-simplex $\sigma ,$ itself a
$(k-1)$-chain, is the boundary $\partial _{k}(\sigma )$ of $\sigma
.$ The boundary of $k$-chain is the sum of the boundaries of the
simplices in the chain. The boundary operator $\partial
_{k}$ is a homomorphism $\partial _{k}:C_{k}\rightarrow C_{k-1}$ and $%
\partial _{k}$'s for $k=0,1,2....$ connect the chain groups into a chain
complex,%
\begin{equation*}
\emptyset \rightarrow C_{n}\overset{\partial _{n}}{\rightarrow }C_{n-1}%
\overset{\partial _{n-1}}{\rightarrow }....\rightarrow C_{1}\overset{%
\partial _{1}}{\rightarrow }C_{0}\overset{\partial _{0}}{\rightarrow }%
\emptyset ,
\end{equation*}
with $\partial _{k}\partial _{k+1}=\emptyset $ for all $k$. The kernel of $%
\partial _{k}$ is the set of $k$-chains with empty boundary while a $k$%
-cycle, denoted by $Z_{k}$, is a $k$-chain in the kernel of
$\partial _{k}.$ The image of $\partial _{k}$ is the set of
$(k-1)$-chains which are
boundaries of $k$-chains with a $k$-boundary, denoted by $B_{k}$, being a $k$%
-chain in the image of $\partial _{k+1}.$%
\begin{eqnarray*}
ker\quad\partial _{k} &=&\left \{ z\in C_{k}:\partial
_{k}(z)=\emptyset
\right \} , \\
im\quad\partial _{k} &=&\left\{ b\in C_{k-1}:\exists b\in C_{k}:%
b=\partial _{k}(z)\right\} .
\end{eqnarray*}%
The collection of $Z_{k}$'s and $B_{k}$'s together with addition
form subgroups of $C_{k}$ while the property $\partial _{k}\partial
_{k+1}=0$ shows that $B_{k}\subseteq Z_{k}\subseteq C_{k},$ i.e.
these groups are nested as illustrated in Figure 2.

\subsection{Homology groups}

The $k$-th homology group is%
\begin{equation}
H_{k}=ker\partial _{k}/im\partial _{k+1}=Z_{k}/B_{k}. \label{h}
\end{equation}

If $z_{1}=z_{2}+B_{k}$, $(z_{1},z_{2}\in Z_{k})$ then the difference
between $z_{1}$ and $z_{2}$ is the boundary and $z_{1}$ and $z_{2}$
are homologous. The $k$-th Betti number of a simplicial complex $K$
is $\beta (H_{k}),$ the rank of the $k$-th homology group, $\beta
_{k}=rank$ $H_{k}.$ or $\beta _{k}=\dim $ $H_{k}$ From expression
(\ref{h}),
\begin{equation}
\beta _{k}=rankH_{k}=rank Z_{k}-rank%
B_{k}.  \label{b}
\end{equation}%
Due to an Alexander Duality property \cite{hatcher}, there is an
intuitive depiction of the first three Betti numbers nicely
explained in \cite{afra}.
Since a non-bounding $0$-cycle represents the set of components of complex $%
K $, there is one basis element per component so that consequently
$\beta _{0}$ represents the number of components of $K$. Hence,
$rank$ $H_{0}=1$ for connected complex $K$ so that the notion of
connectivity is reflected in $H_{0}$. A non-bounding $1$-cycle
represents a collection of non-contractible closed curves in $K$, or
based on duality property, a set of tunnels formed by $K$. Each
tunnel can be represented as a sum of tunnels from the basis so that
$\beta _{1}$ represents the dimension of the basis for the tunnels.
These tunnels may be perceived as forming graph with cycles
\cite{afra}. A $2 $-cycle which itself is not a boundary represents
the set of non-contractable closed surfaces in $K$, or based on
duality principle, a
set of voids which exist in the complement of the simplicial complex , i.e. $%
\mathbb{R}^{3}-K.$ The dimension of the basis for voids, equal to
the number of voids is represented by $\beta _{2}.$

\section{Construction of Simplicial Complexes from Graphs}

Simplicial complexes may be constructed from undirected or directed
graphs (digraphs) in several different ways. Here we only consider
two of them: the
neighborhood complex and the clique complex. The neighborhood complex \emph{N%
}$(G)$ is constructed from the graph $G$, with vertices
$\{v_{1},...,v_{n}\}$ in such a way that for each vertex $v$ of $G$
there is a simplex containing
the vertex $v$, along with all vertices $w$ corresponding to directed edges $%
v\rightarrow w.$ The neighborhood complex is obtained by including
all faces of those simplices and in terms of matrix representation,
the incidence matrix is obtained from the adjacency matrix of $G$ by
increasing all diagonal entries by $1$. An example of the
construction of a neighborhood complex is represented in Fig. $3$.
The clique complex $C(G)$ has the complete subgraphs as simplices
and the vertices of $G$ as its vertices so that it is essentially
the complete subgraph complex. The maximal simplices are given by
the collection of vertices that make up the cliques of $G$. In
literature, a clique complex is also referred to as flag complex. An
example of a clique complex is presented in Fig. 4.

These two methods are not the only ones that may be used for
constructing simplicial complexes from graphs. Actually, any
property of the graph $G$ that is preserved under deletion of
vertices or edges may be used for construction purposes. A detailed
account of the methods for obtaining simplicial complexes from
graphs, among many other issues related to the
relationship between graphs and simplicial complexes, may be found in \cite%
{johnss}.

\section{Persistent homology}

\subsection{Filtration}

The basic aim of persistent homology \cite{edels} is to measure
life-time of certain topological properties of a simplicial complex
when simplices are added to the complex or removed from it. Usually
the evolution of the complex considers its creation starting from
the empty set, hence the assumption is that simplices are added to
the complex (corresponding to the growing network). The sequence of
subcomplexes constructed in the process is known as filtration. In
more formal terms the filtration of the simplicial
complex $K$ is a sequence of complexes $K_{i}$, such that:%
\begin{equation*}
\emptyset =K_{0}\subset K_{1}\subset ...\subset K_{n}=K.
\end{equation*}%
The simplices in $K$ are indexed by their rank in a filtration
sequence and each prefix of the sequence is a subcomplex. Two
filtration constructions are usually considered when the history of
the complex is studied. The first one is formed when at each stage
of the filtration only one simplex is added (i.e. $K_{i}/K_{i-1}$
consists of one simplex $\sigma _{i}$ for each $i$). In the second
case a simplex $\sigma _{i}$ is added to the sequence, say to
subcomplex $K_{j}$, when all its faces are already parts of some $K_{i}$ $%
(i\leq j).$ Hence, the second case does not require only one simplex
to be added at each stage of filtration. These two filtrations
contain complete orderings of its simplices and Figure 5 illustrates
the two progressive sequences. Naturally, other filtrations may also
be applied in practice including "irregular" ones when simplices are
removed or disappear in the sequence. For these filtrations the main
aspect of change is not only growth but decrease as well.

\subsection{Algebraic formulation of persistent homology}

Following the expositions in the pioneering paper on persistent
homology \cite{edels} and in reference \cite{afra} we give here some
basic notions and concepts. Persistence is defined in conjunction
with cycle and boundary groups of complexes in filtration i.e. with
respect to homology groups and associated Betti numbers. Since
homology captures equivalent classes of cycles by factoring out the
boundary cycles, the focus is on the count of non-bounding cycles
whose life-span lasts beyond a chosen threshold (say represented by
number $p$ of next complexes in the filtration sequence) and which
determine persistent or long lasting topological properties of the
complex. These cycles persist through $p$ phases of the sequence,
hence they are important. In a complementary manner our interest
also lies in cycles with short life-spans which convert to
boundaries during filtration. Algebraically, it is relatively simple
to perform the count of persistent non-bounding cycles. Let
$Z_{k}^{l}$ and $B_{k}^{l}$ represent the $k$-th cycle group and the
$k$-th boundary group, respectively, of the $l$-th complex $K^{l}$
in filtration sequence. In order to obtain the long-lasting
non-bounding cycles, the $k$-th cycle group is factored by the
$k$-th boundary group of the $K^{l+p}$ complex, $p$ complexes later
in the
filtration sequence. Formally, the $p$-persistent $k$-th homology group of $%
K^{l}$ is
\begin{equation}
H_{k}^{l,p}=Z_{k}^{l}/(B_{k}^{l+p}\cap Z_{k}^{l}).  \label{ph}
\end{equation}%
Clearly $B_{k}^{l+p}\cap Z_{k}^{l}$ is a group itself being an
intersection
of two subgroups of $C_{k}^{l+p}.$ The $p$-persistent $k$-th Betti number, $%
\beta _{k}^{l+p}$ of the $l$-th complex $K^{l}$ in filtration is the
rank of
$H_{k}^{l,p}:$%
\begin{equation*}
\beta _{k}^{l+p}=rank H_{k}^{l,p}.
\end{equation*}%
Hence, $\beta _{k}^{l+p}$ counts homological classes in the complex
$K^{p}$ which were created during filtration in the complex $K^{l}$
or earlier. There is a Betti number for each dimension $p$ and for
every pair of indices $(k,p),$ $0\leq k\leq p\leq n.$ To get a more
intuitive illustration of persistence concept let us consider a
non-bounding $k$-cycle created at time (step) $i$ as a consequence
of the appearance of the simplex $\sigma $ in the complex so that
the homology class of $z$ is an element of $H_{k}^{i}$, i.e. $[z]\in
H_{k}^{i}.$ The simplex $\sigma $ will be labelled as a $creator $
simplex, or $\sigma ^{+}($positive simplex). Consider the appearance
of another simplex $\tau $ at time $j\geq i$ which turns a cycle
$z^{\prime }$ in $[z]$ into a boundary, so that $z^{\prime }\in
B_{k}^{j}.$ This causes the decrease of the rank of the homology
group since the class $[z]$ is joined with the older class of
cycles. The simplex $\tau $ will be labelled as an $annihilator$
simplex, $\tau ^{-}$(negative simplex) since it annihilates both
$z^{\prime }$ and $[z]$. The persistence of $z$ and its homology
class $[z]$ is then $j-i-1$. As $p$ increases by one step (assuming
full ordering of simplices), persistence of all non-bounding cycles
is decreased by one so that while $p$ increases, negative simplices
cancel positive ones which appeared earlier in the filtration. For
$p$ large (long enough), the topological noise may be removed from
pertinent information about homology groups and Betti numbers.

\subsection{Computational remarks}

High quality algorithms exists for the computation of homology
groups with respect to various applications. A comprehensive
introduction to the subject of computational homology is
\cite{mishaikow} (and the associated COmputational HOmology project
CHOMP \cite{chomp}) while the algorithms for persistent homology are
given in \cite{edels} and \cite{carlsson}. Various practical issues
related to these algorithms and computational methods for evaluating
Betti numbers are covered in \cite{horak}. The Matlab-based program
"Plex" \cite{silva}, designed for the homology analysis of point
cloud data set converted into a global simplicial complex (Rips,
\v{C}ech or other) offers a wealth of possibilities for various
operations with simplicial complexes and simplicial homology
calculations. Also, a package "Simplicial Homology" \cite{dumas},
requiring "GAP" \cite{gap}, a system for computational discrete
algebra with special emphasis on computational group theory,
provides numerous functionalities related to simplicial homology.
Both Plex and Simplicial Homology have been extensively used in our
calculations presented here.

\section{Visualization of persistence homology: Barcodes}

Since persistent homology represents an algebraic invariant that
detects the birth and death of each topological feature as the
complex evolves in time, it is advantageous to encode the persistent
homology in the form of a parametrized version of the rank of
homology group i.e. its Betti number \cite{ghrist}. One possible
choice for the parameter is time as it encompasses both the case of
simplex growth (addition of simplex or simplices) and of its
decrease (removal of simplex or simplices). The other choice is to
use intervals whose endpoints are filtration complexes essentially
representing the filtered simplicial complex at times when the
addition (or removal) of simplices takes place. During its temporal
existence, each topological attribute plays a part in the formation
of some Betti number and our interest lies in those properties with
long lifetimes (persistent properties). The parameter intervals
represent lifetimes of various stages of filtration and they may be
represented on the horizontal axis while arbitrary ordered homology
generators $H_{k}$ may be represented on the vertical axis. Figure 6
shows an example of filtration and the
barcode for $H_{k}(k=1,2,3)$.

The rank of persistent homology group $%
H_{k}^{i\rightarrow j}$ equals the number of intervals in the
barcode of homology group $H_{k}$ within the limits of the
corresponding parameter
range or lifetime $[i,j]$. Here $i$ and $j$ may denote filtration times $%
t_{i}$ and $t_{j}$ or filtration complexes $K_{i}(t_{i})$ and $K_{j}(t_{j})$%
. Clearly, barcodes do not provide information on delicate structure
of the
homology however the information about persistent parametrized rank (since $%
\beta _{k}=rank$ $H_{k}$ a barcode reflects the persistent
properties of Betti numbers) enables clear distinction between
topological noise and topological "signal".

\section{Persistent homology of random networks}

For the purpose of illustrating persistent homology we first
consider random (Erd\"{o}s-R\'{e}nyi) networks $G(n,p)$ for which
the number of nodes, $n,$ is fixed and with each link inserted with
the same probability $p.$ As is well known, a random network has a
characteristic scale in its node connectivity reflected by the peak
of the distribution which corresponds to the number of nodes with
the average number of links. We have constructed the clique complex
$C(G)$ of a random network so that the obtained complex is a random
simplicial complex $C(G(n,p)).$ The filtration $F$ of the complex is
\begin{equation}
F=\{ K_{0},K_{1},...K_{n} \} $ such that $K_{0}\subset K_{1}\subset
...\subset K_{n}=K.  \label{f1}
\end{equation}%
Then the $i$-th complex in the filtration is given by%
\begin{equation}
K_{i}=\sum_{j=1}^{i}S_{j},  \label{f2}
\end{equation}%
where $S_{j}$ is the $j$-th skeleton of the clique complex (the set
of simplices of dimension less or equal to $j$).

The random network considered consists of $2000$ nodes with the
probability of two nodes having a link equal to $p=0.005$. The
corresponding barcode is presented in Fig. 7. Due to sparsity of the
network the filtration steps are limited to complexes of dimension
$3$. It is evident that persistent $H_{0}$ has Betti number $\beta
_{0}$ $=1$ corresponding to one line that persists through all
stages of filtration. Since the zero dimensional homology measures
the connectivity of the underlying graph the graph is always
connected and this property remains for arbitrary choice of $p$
$(0\leq
p\leq 1)$ or $n$, as one would expect. In addition $\beta _{1}=rank$ $%
H_{1}=7847$ while $\beta _{2}=rank$ $H_{2}=0.$ The maximal rank of
persistent homology of this random network is $1$. However, due to
the short lifetime of $H_{1}$ through only two filtrations, it may
be inferred that the content of topological noise dominates the
network for this choice of parameters $p$ and $n$. The same results,
from the aspect of persistence,
are obtained for the neighborhood complex $(\beta _{0}=1,$ $\beta _{1}=$ $%
13503,$ $\beta _{2}=0).$ Increasing the probability $p$ or the
number of nodes $n$ leads to occurrence of higher dimensional
homology groups which though appear only as noise as illustrated in
Fig. 8 for the case of $p=0.02$ and $n=2000$. There is an interval
outside which homology vanishes, and
inside which only lowest ranked homologies persist, i.e. $H_{0}$ and $H_{1}$%
. This conlcusion is in agreement with recent theoretical studies on
clique and neighborhood complexes of random graphs \cite{kahle1},
\cite{kahle2}.

\section{Persistent homology of a network with exponential connectivity
distribution}

In order to analyze the emergence of self-similar properties in a
complex network, an e-mail network was studied in \cite{arenas}.
Each e-mail address in this network represents a node and links
between nodes indicate e-mail communication between them. After
removal of bulk e-mails, the connectivity
distribution of this network is exponential, $P(k)\exp (-k/k^{\ast })$ for $%
k\geq 2$ and with $k^{\ast }=9.2.$ The number of nodes (e-mail
users) is 1700. Calculations were performed using both the clique
and the neighborhood complex and both showed consistent persistency
property. The corresponding persistency barcode is presented in Fig.
9 in which the rank of the homology group equals the number of
intervals in the barcode intersecting the dashed line which
corresponds to the filtration stage. The first three homology
groups, i.e. $H_{0}$, $H_{1}$ and $H_{2},$ have long lived
generators while higher dimensional homology groups appear only as
topological noise. Although random networks analyzed earlier and the
e-mail network have comparable number of nodes, the number of higher
dimensional homology groups is considerably larger in the latter
case. This is the consequence of\ an
internal organization of an e-mail network into a number of communities \cite%
{arenas} which is an essential prerequisite for emergence of higher
dimensional complete graphs. Clearly, no such organizational
principle exists in random networks (random simplicial complexes)
and $1$-cycles dominate the complex. The fact that homology groups
of dimension higher than $2$ have short lifetimes indicates that
communications among certain groups of e-mail users may not exist
for a certain time during the growth of the network however these
communication channels are reestablished at later stages of the
network evolution.

\section{Persistent homology of scale-free networks}

Among scale-free networks we consider scale-free models with modular
structure developed recently \cite{bosa}. The model including
preferential-attachement and preferential-rewiring during the graph
growth is generalized so that new modules are allowed to start
growing with finite probability. The structural properties of
modular networks are controlled by three parameters: the average
connectivity $M$, the probability of the emergence of a new modul
$P_{0}$ and the attractiveness of the node $\alpha . $ By varying
these parameters the internal structure of modules and the network
connecting various modules is kept under control. Detailed
explanation of the role of each of these parameters in the control
process are discussed in \cite{bosa}. Here we consider the
persistent homology of
three scale-free networks developed using three diferent sets of parameters $%
(M,$ $P_{0},$ $\alpha ),$ chosen as paradigmatic for the type of
network considered. The results for both clique and neighborhood
complexes were constructed and since the results do not differ for
the two cases the
presented ones are obtained from the clique complex filtration \ref{f1} and %
\ref{f2}. All networks were generated with 1000 nodes.

\subsection{Case 1. Clustered modular network}

The average connectivity (number of links per node) is $M=5$. The
network has $1000$ nodes and $7$ modules so that $P_{0}=0.007.$ The
attractiveness of the node is $\alpha =0.6$ which enables stronger
clustering effect, hence the label "clustered modular network". The
corresponding barcodes are presented in Fig. 10.\ There are unique
persistent generators for $H_{0}$ and $H_{1}$ while for $H_{2}$
there are $2$ persistent generators. $H_{3}$ also has a persistent
generator which starts at stage $2$ of filtration. It is interesting
that once the homology is generated at later stages of filtration it
remains persistent for all $H_{i}$ $(i=1,2,3)$ as indicated by
arrows. One aspect of existence of persistent homology groups is
robustness of the complex (network) with respect to addition or
reduction of simplices (nodes). The fact that four homology groups
show persistence is a clear sign of robustness. Moreover,
practically there is no topological noise in this case.

\subsection{Case 2. Clustered non-modular network}

The parameters for this type of network are $M=5,$ $P_{0}=0$ (no
modules) and $\alpha =0.6$ (strong clustering). The persistence
barcodes for this network are presented in Fig. 11. The most
striking feature of these topological persistency representations is
the existence of $H_{4}.$ Another striking feature is that $H_{3}$
does not exist for this particular value of clustering parameter
$\alpha $ showing that higher ranked persistency generators may not
be distributed continuously across dimensions. There are four
generators for $H_{4}$ however they persist through five stages of
filtration and there are several more generators with shorter
lifetime some of which may be considered as topological noise, such
as the ones whose lifetime is one or two filtration phases. The fact
that $H_{3}$ generators
do not exist shows that for this choice of parameters there are no $3$%
-dimensional non-bounding cycles in the complex.

\subsection{Case 3. Modular non-clustered network}

The average connectivity is $M=5$. Modular probability is
$P_{0}=0.007$ and clustering coefficient $\alpha =1.0$ so that there
is only one link between each of the modules and effectively there
is non clustering. The corresponding barcodes are shown in Fig. 12.
There is only one generator for $H_{0}.$ For $H_{1}$ there is a
unique generator persistent from the beginning of filtration however
there are several generators which persist while occurring with the
slight delay in filtration sequence. The maximal persistent homology
rank is $2$ and $H_{2}$ has relatively long lived generators with a
slight noise. Of the three cases considered this one has
the smallest number of persistent homology groups, namely three ($H_{0}$, $%
H_{1}$ and $H_{2}$), and also the smallest number of generators for
the homology group $H_{2}$.

\subsection{Remarks on persistent homology of scale-free networks}

Since both clustered modular and clustered non-modular networks have
higher ranked persistent homology ($H_{3}$ and $H_{4}$ respectively)
then the non-clustered modular network ($H_{2}$), it is clear that
clustered networks are more robust with respect to addition
(removal) of nodes (simplices). Moreover, clustering property is
more important for robustness then modularity as may be also
inferred by comparison with the e-mail network discussed in Sec. 7
which also shows modular structure. The fact that only $1
$-dimensional and $2$-dimensional cycles (voids) are persistently
missing in
non-clustered simplices with respect to additional lack of $3$ and $4$%
-dimensional cycles in modular simplices may convey important
information depending upon the context of the analysis and types of
networks under study. In general the persistence of $n$-th homology
generators ($n$-th Betti numbers) means that somewhere in the
complex $n$-th dimensional subcomplex is missing through all stages
of complex growth or reduction. In other words an $n$-dimensional
object formed by simplices of dimension at most $n$ is absent from
the complex. This property may be translated to the "network
language" in terms of connectivity relations which depend on the
context. In simplified terms, for example for $n=2$ the network
lacks in dyadic (binary) relations; for $n=2$ there are no triadic
(ternary) relations and so on where $n$-adic relations should be
regarded not only as the set of its node-to-node relations but in
their relational entirety. As an example, a face of a triangle
represents a relational entirety (essentially a relationship of
higher order) of a three node relation.

\section{Summary and concluding remarks}

Construction of simplicial complexes from graphs (networks) creates
a topological setting which offers flexible tools for gauging
various topological attributes. Here our interest lies in detection
of long lived homology groups of a simplicial complex (network)
during the course of its history which includes both addition and
removal of simplices (nodes). The method relies on visual approach
of recognizing persistent features in the form of a barcode which
may be regarded as the persistence analogue of a Betti number. The
results show distinct persistency attributes for random networks,
networks with exponential degree distributions and for scale-free
networks. Persistency includes the two lowest dimensional homology groups $%
H_{0}$ and $H_{1}$ for random networks. For the case of neworks with
exponential degree distribution persistency includes $H_{0},$ $H_{1},$ and $%
H_{2}$ while for scale-free networks persistent homology groups are
$H_{0},$ $H_{1},$ $H_{2}$, $H_{3}$ and even $H_{4}$. An obvious
consequence of persistency is that it gives important information
about robust quality of the network so that scale-free networks,
especially the ones with clustering properties, exhibit the highest
topological resilience to change in the form of addition or removal
of the nodes. However persistence of certain topological attributes
assumes also long lived defficiency in certain topological forms in
simplicial complexes corresponding to defficiency of certain
relations in networks. In order to reveal more about the sense of
balance between these two properties we will use more subtle
topological methods in our future work.

\ack{The authors gratefully acknowledge the support of the Ministry
of Science of the Republic of Serbia through research grant OI
144022.}

\eject

\begin{center}
\includegraphics[width=100mm]{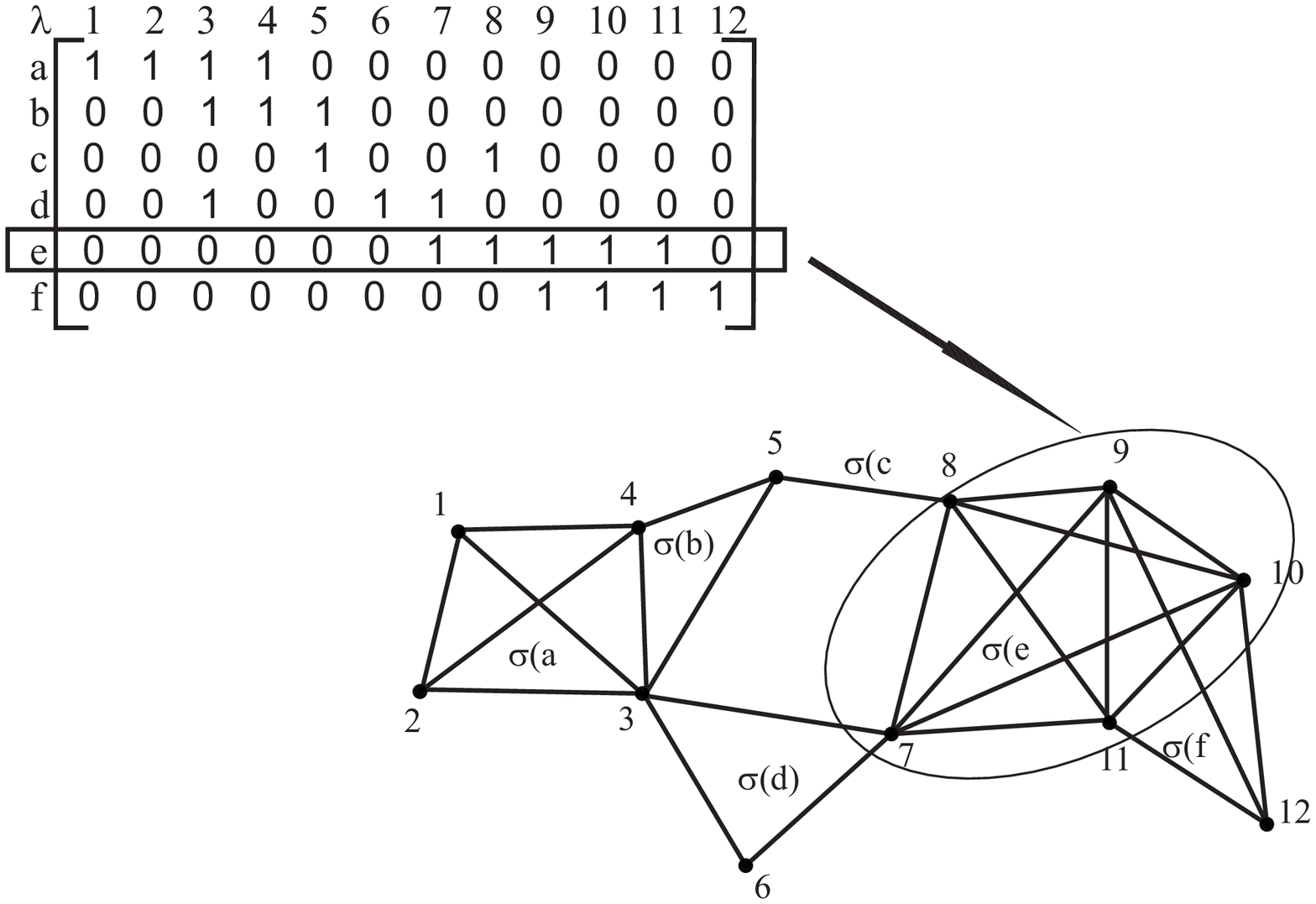}

Fig. 1 An example of a simplicial complex and its incidence matrix
representation. Columns are labeled by its vertices and rows are
labeled by its simplices.
\end{center}
\bigskip

\begin{center}
\includegraphics[width=100mm]{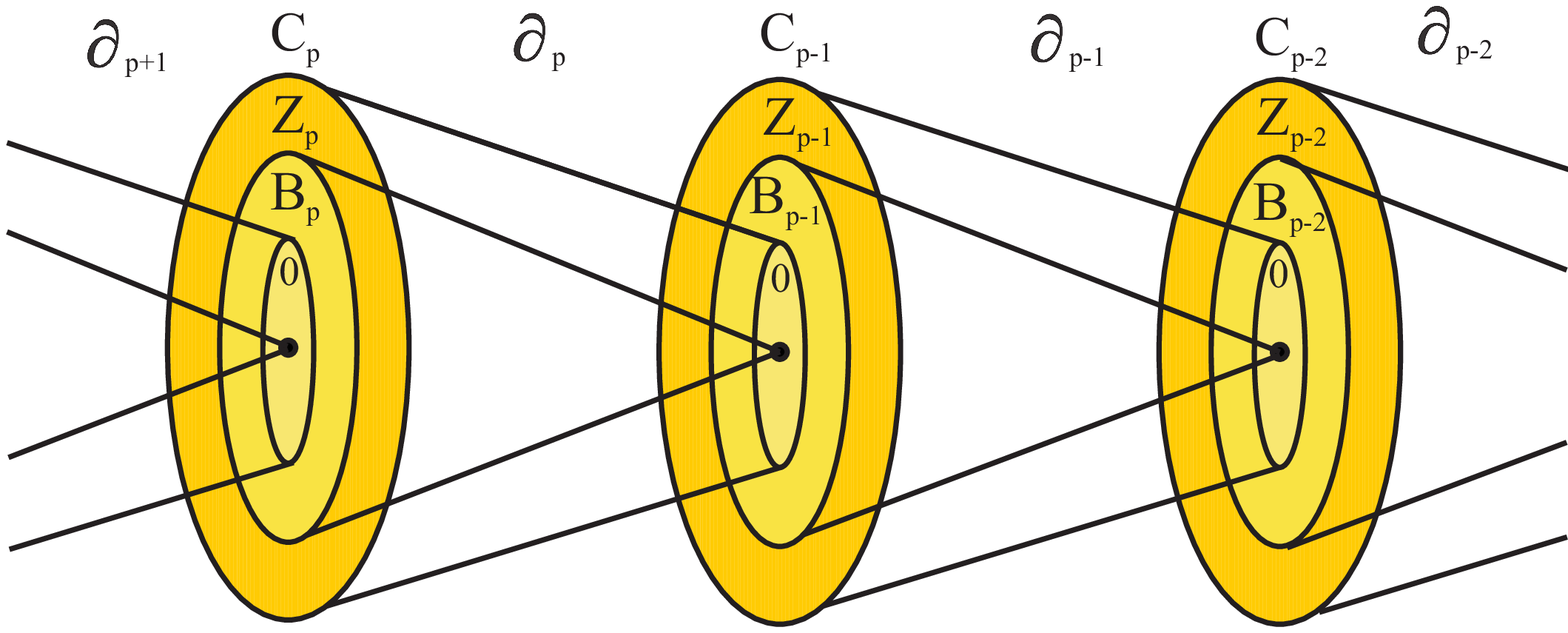}

Fig.2 Chain, cycle and boundary groups and their mappings under
boundary operators.
\end{center}

\bigskip

\begin{center}
\includegraphics[width=100mm]{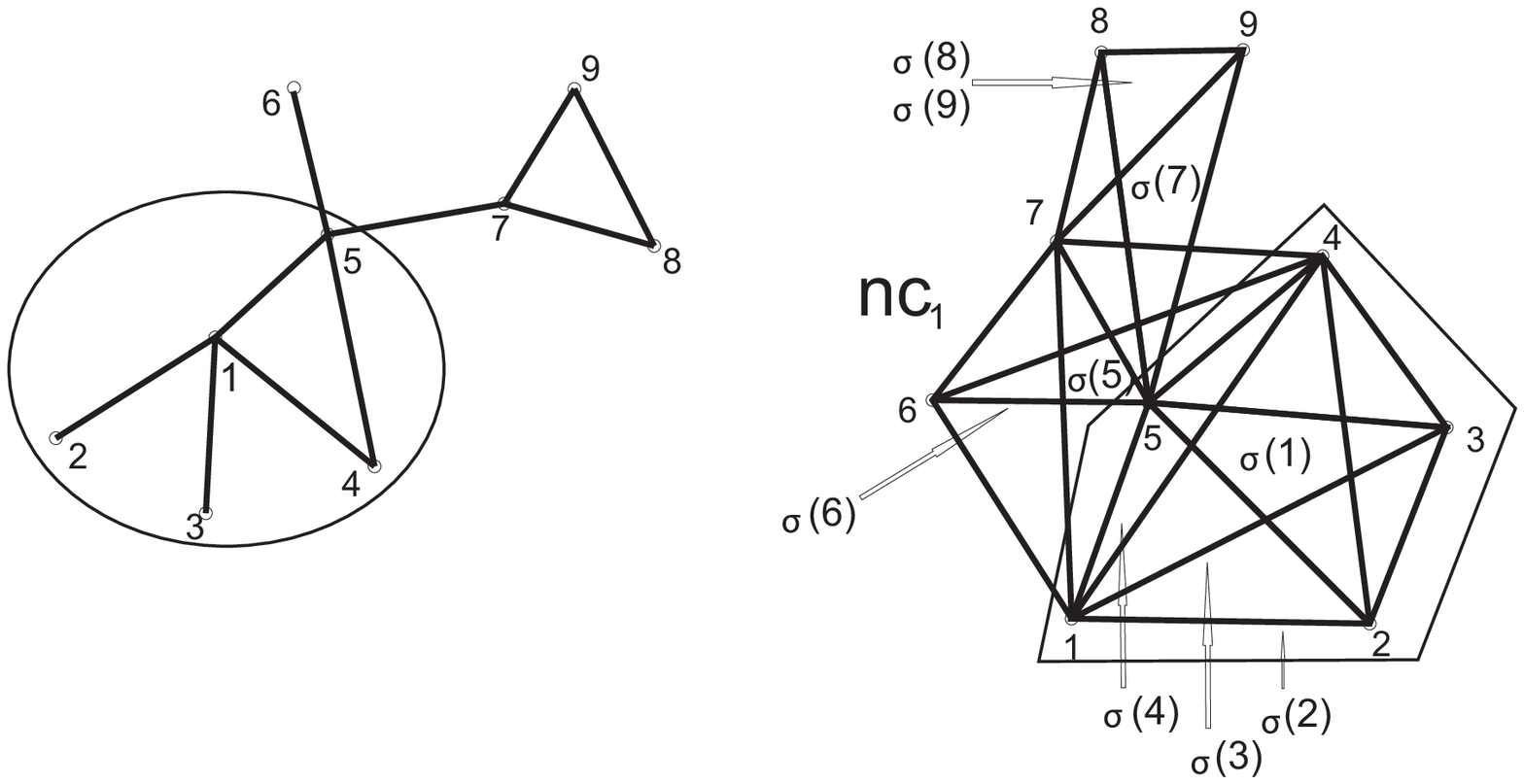}

Fig. 3 A graph and its associated neighborhood complex. Simplices
are labeled as $\sigma (i)$, where $i$ denotes the vertex whose
neighbors define the simplex.
\end{center}

\bigskip

\begin{center}
\includegraphics[width=70mm]{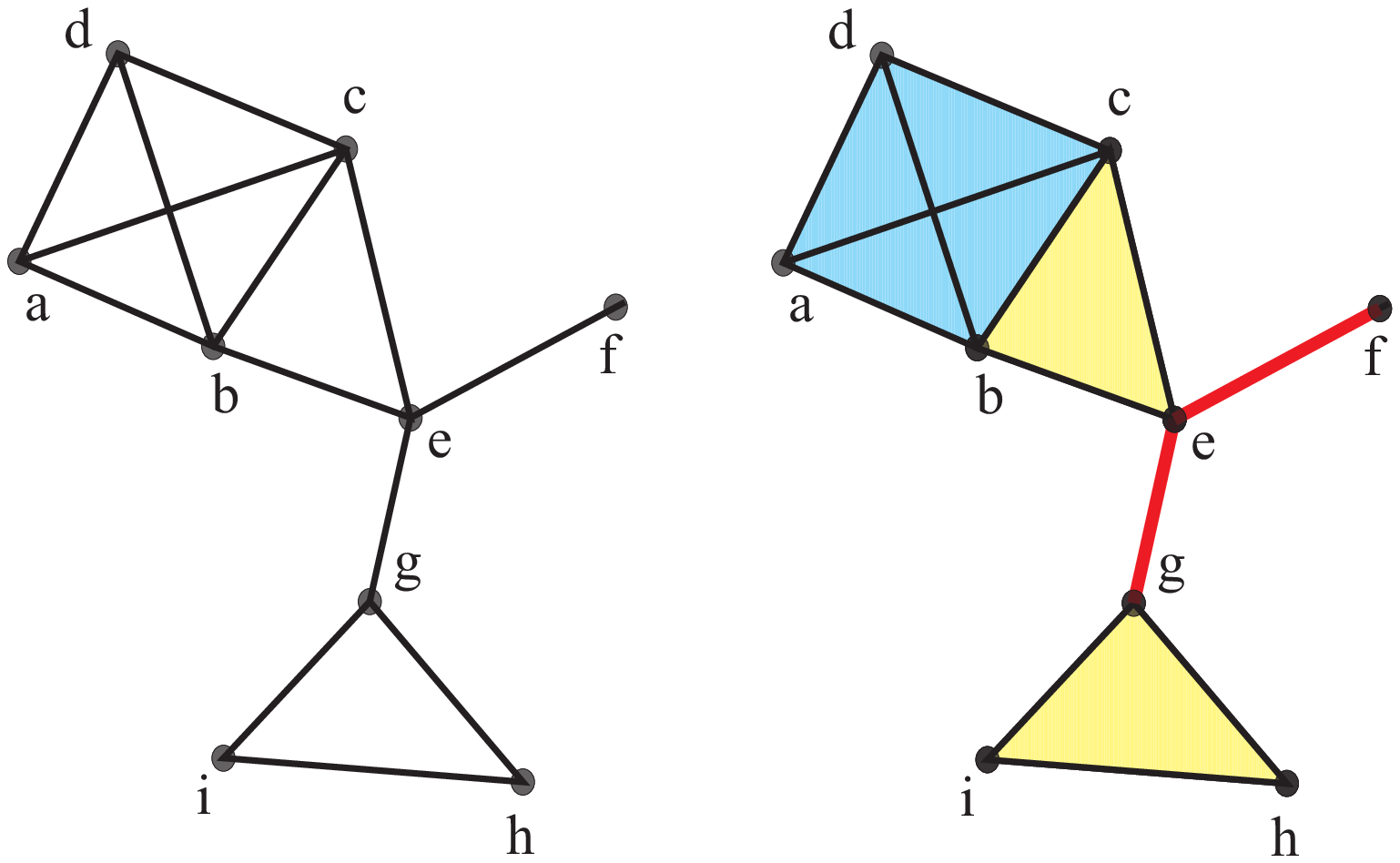}

Fig. 4 An example of a graph and its associated clique-complex.
\end{center}

\bigskip

\begin{center}
\includegraphics[width=120mm]{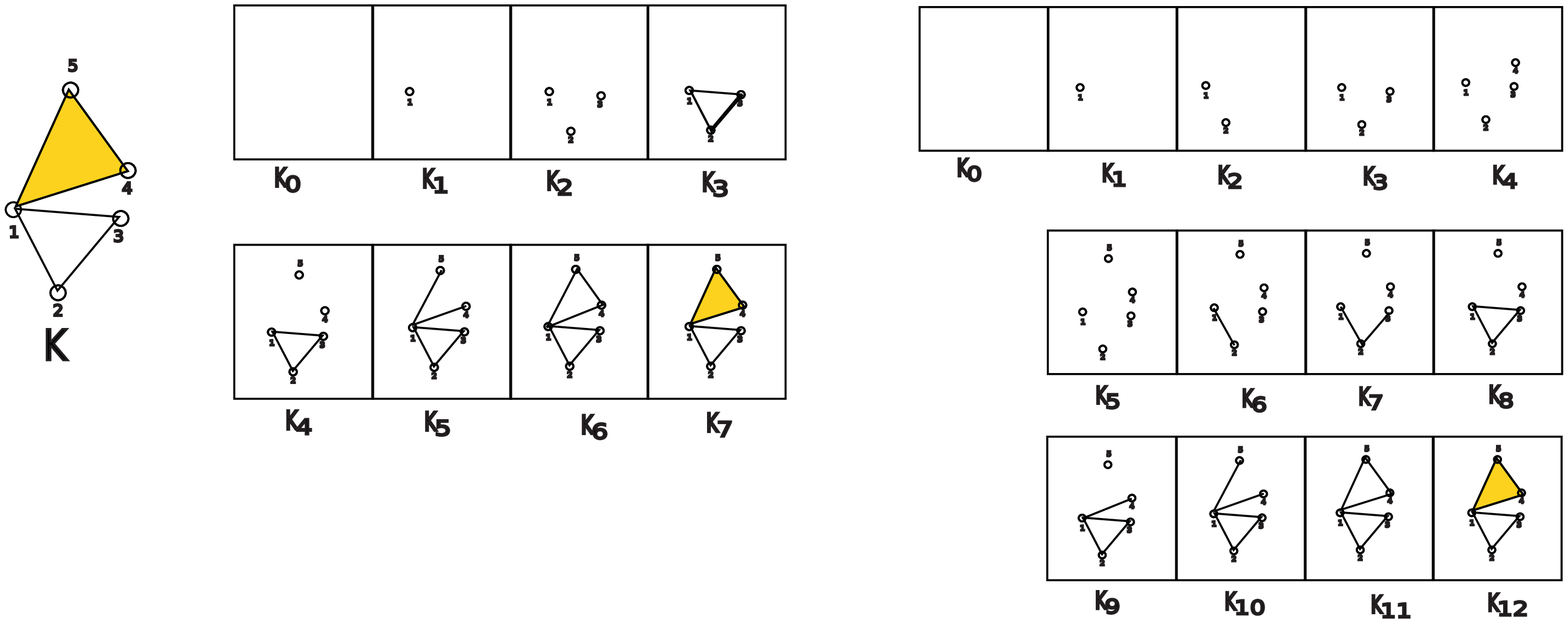}

Fig. 5 The simplicial complex \thinspace $K$ and its two
filtrations. In the filtration on the right one simplex is added at
each phase of the sequence.
\end{center}

\bigskip

\eject
\begin{center}
\includegraphics[width=100mm]{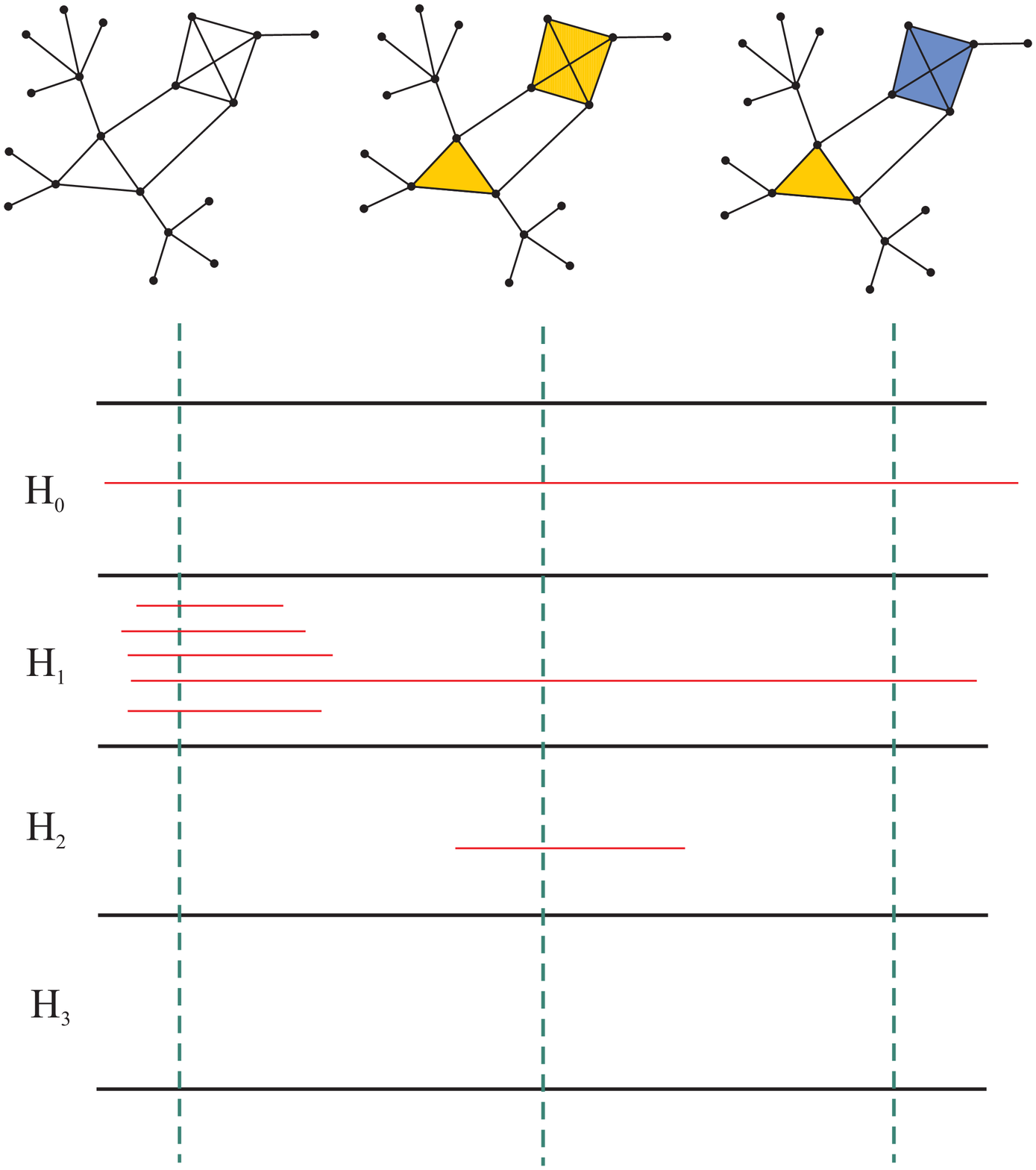}

Fig. 6 An example of the barcode for an arbitrary simplicial
complex. The rank of $H_{k}$ equals the number of parameter
intervals traversed by the red barcode line.
\end{center}

\bigskip

\begin{center}
\includegraphics[width=100mm]{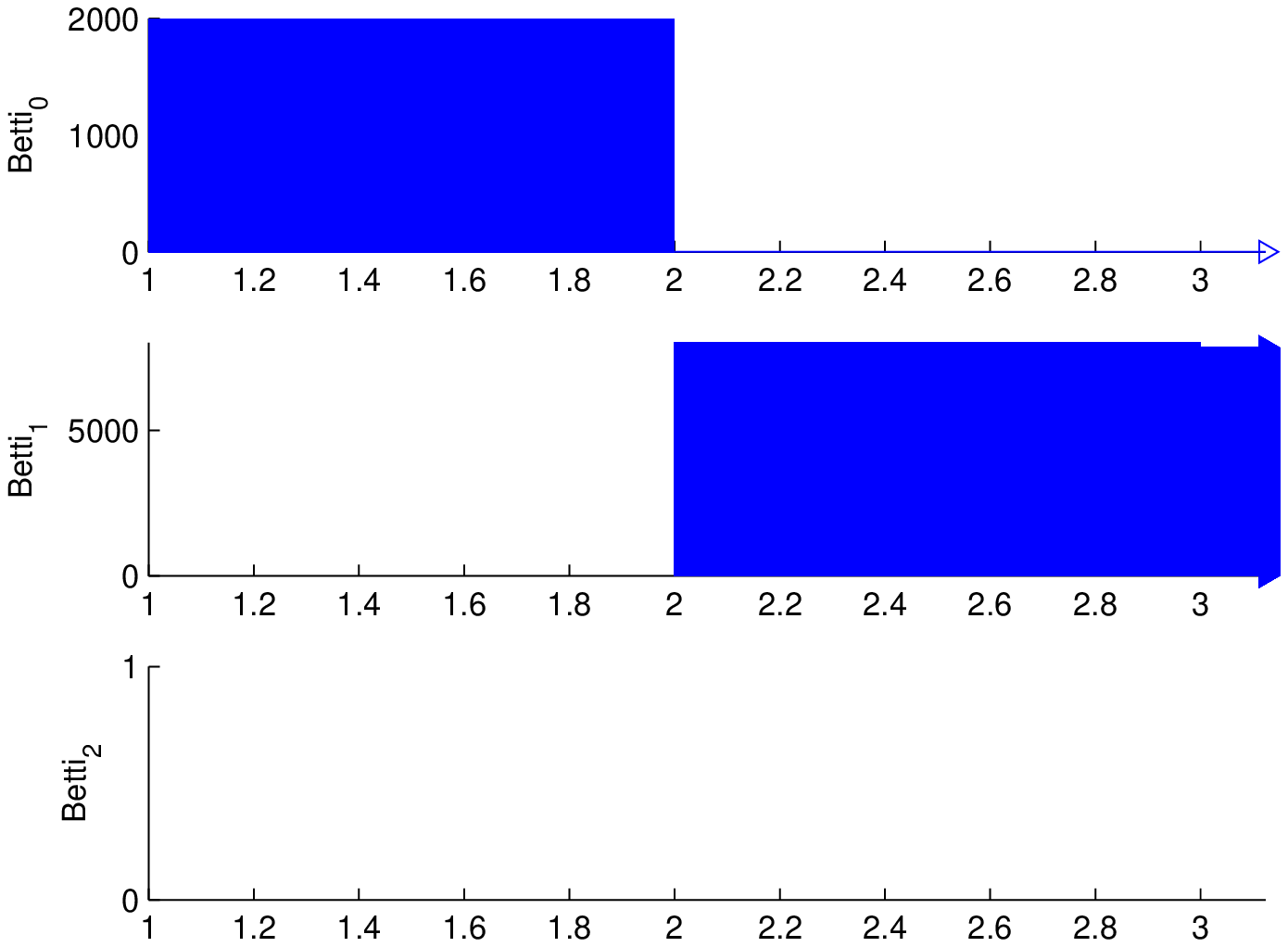}

Fig 7. Barcode of $G(n,p)$ random network with $n=2000$ and
$p=0.005$.
\end{center}

\bigskip

\begin{center}
\includegraphics[width=100mm]{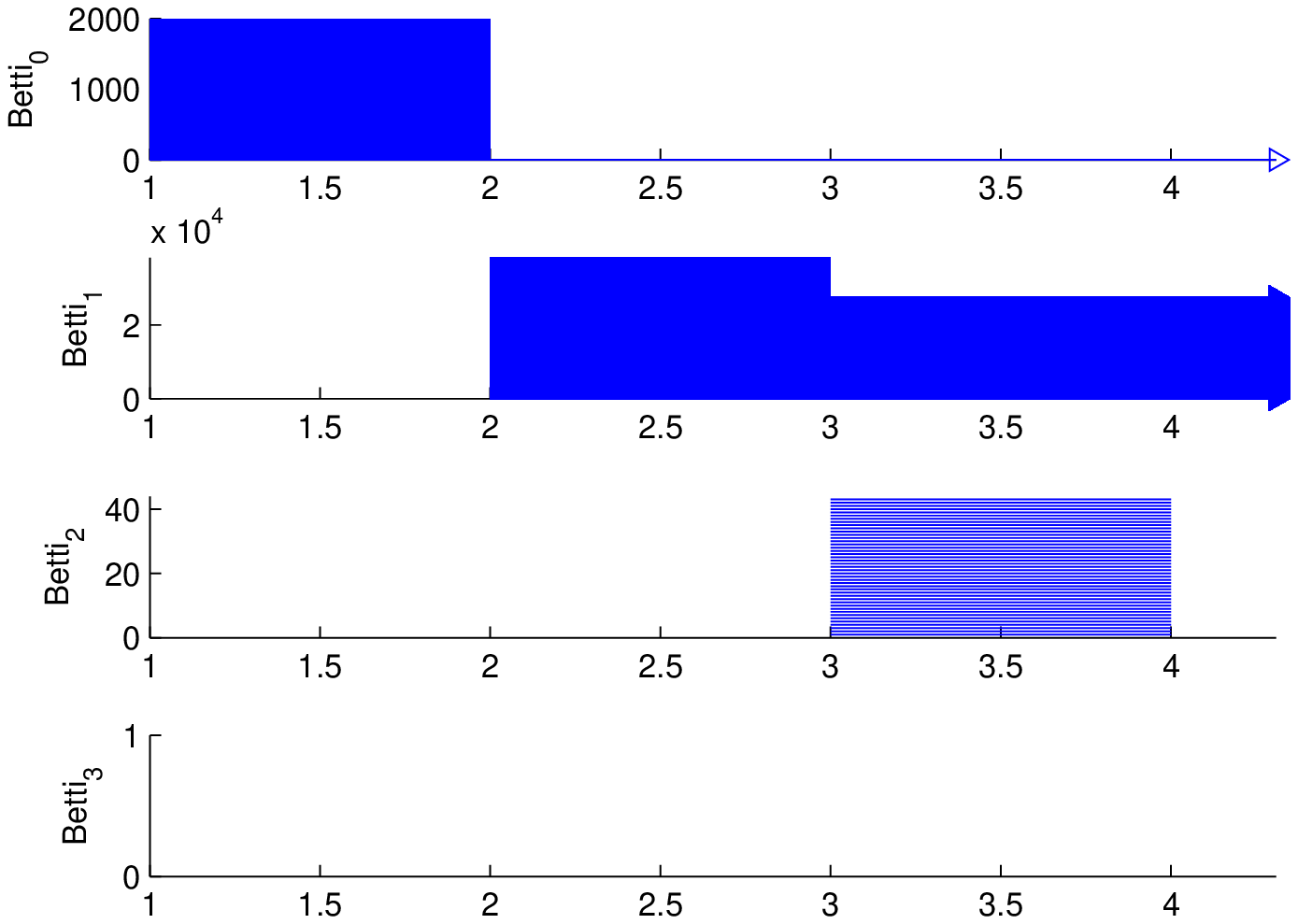}

Fig. 8 Barcode of $G(n,p)$ random network with $n=2000$ and
$p=0.02$.
\end{center}

\bigskip

\begin{center}
\includegraphics[width=100mm]{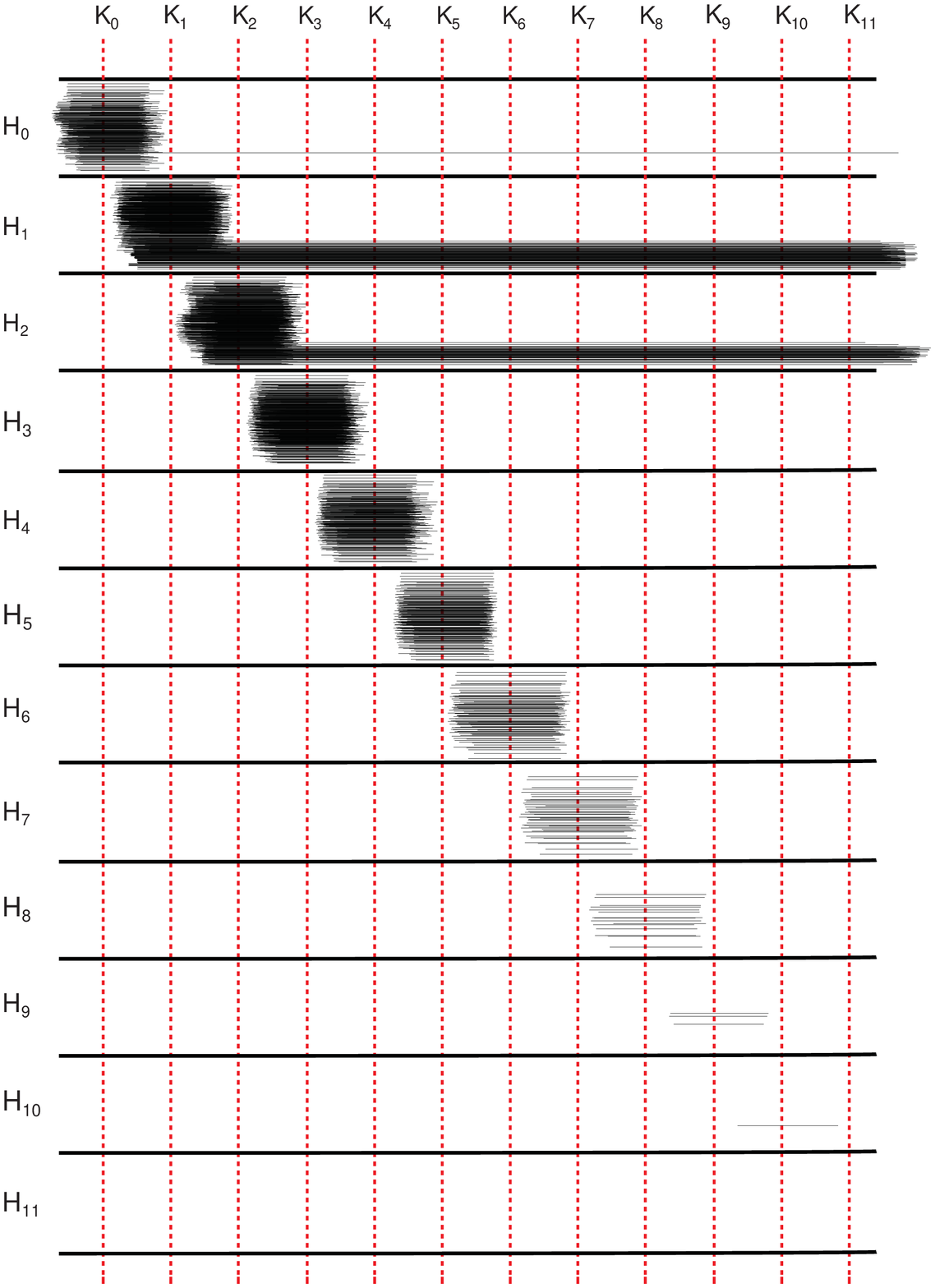}

Fig. 9 Barcode of the e-mail network with exponential degree
distribution.
\end{center}

\bigskip

\begin{center}
\includegraphics[width=110mm]{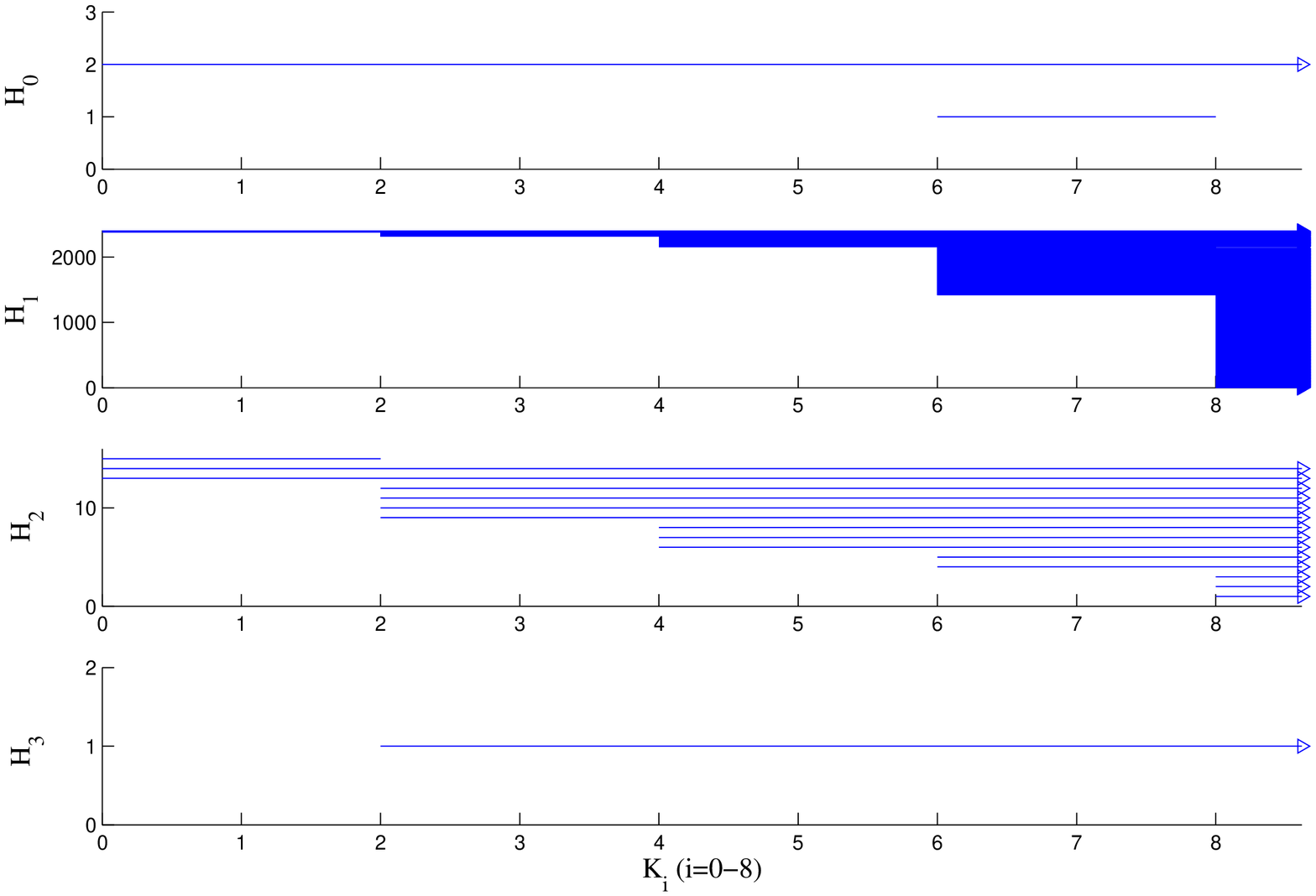}

Fig. 10 Barcode of the clustered modular network. Persistence of all
homology groups may be easily noticed.
\end{center}

\bigskip

\begin{center}
\includegraphics[width=110mm]{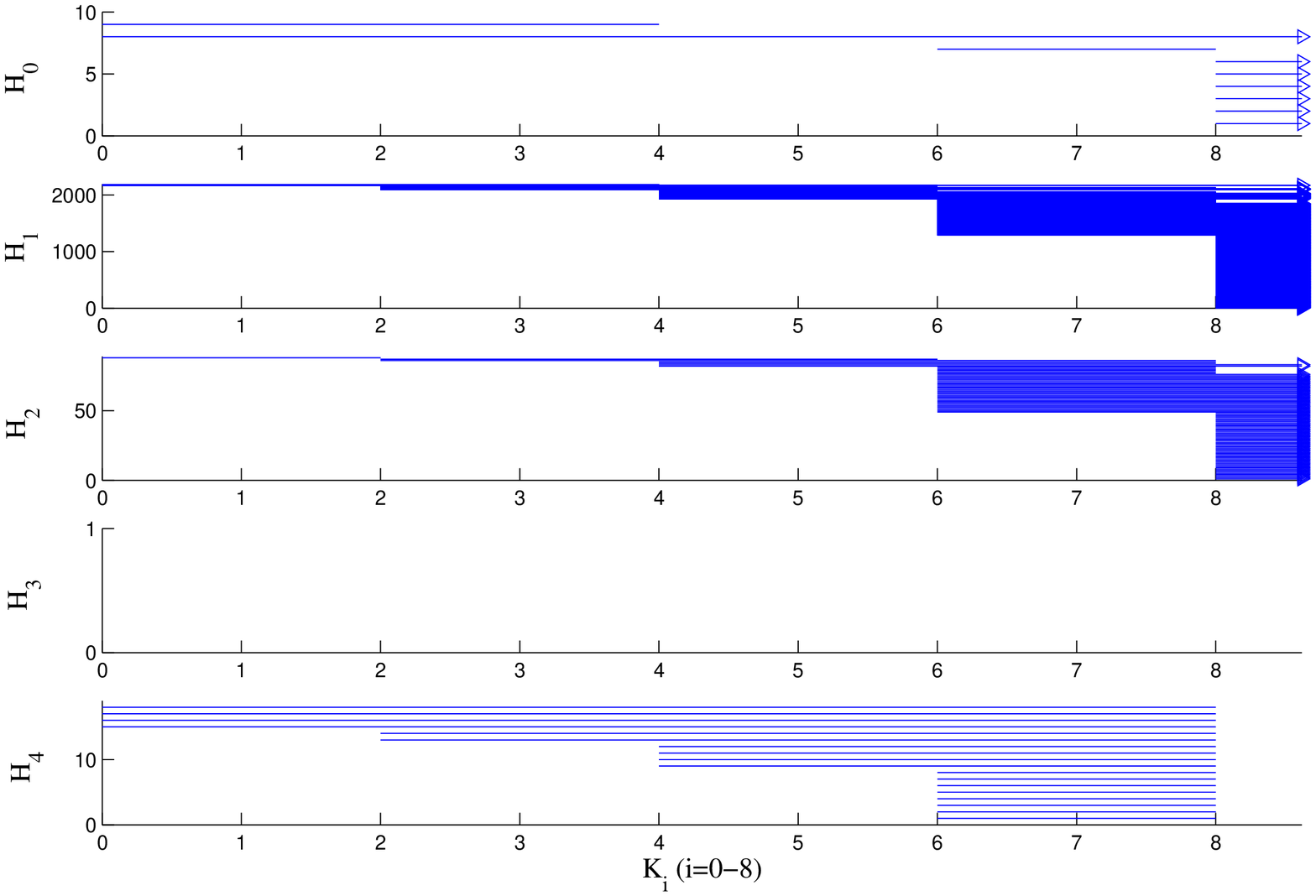}

Fig. 11 Barcode of the clustered non-modular network. The rank of
$H_{3}$ is
equal to zero, however the rank of the highest persistent homology group is $%
4$.
\end{center}

\bigskip

\begin{center}
\includegraphics[width=110mm]{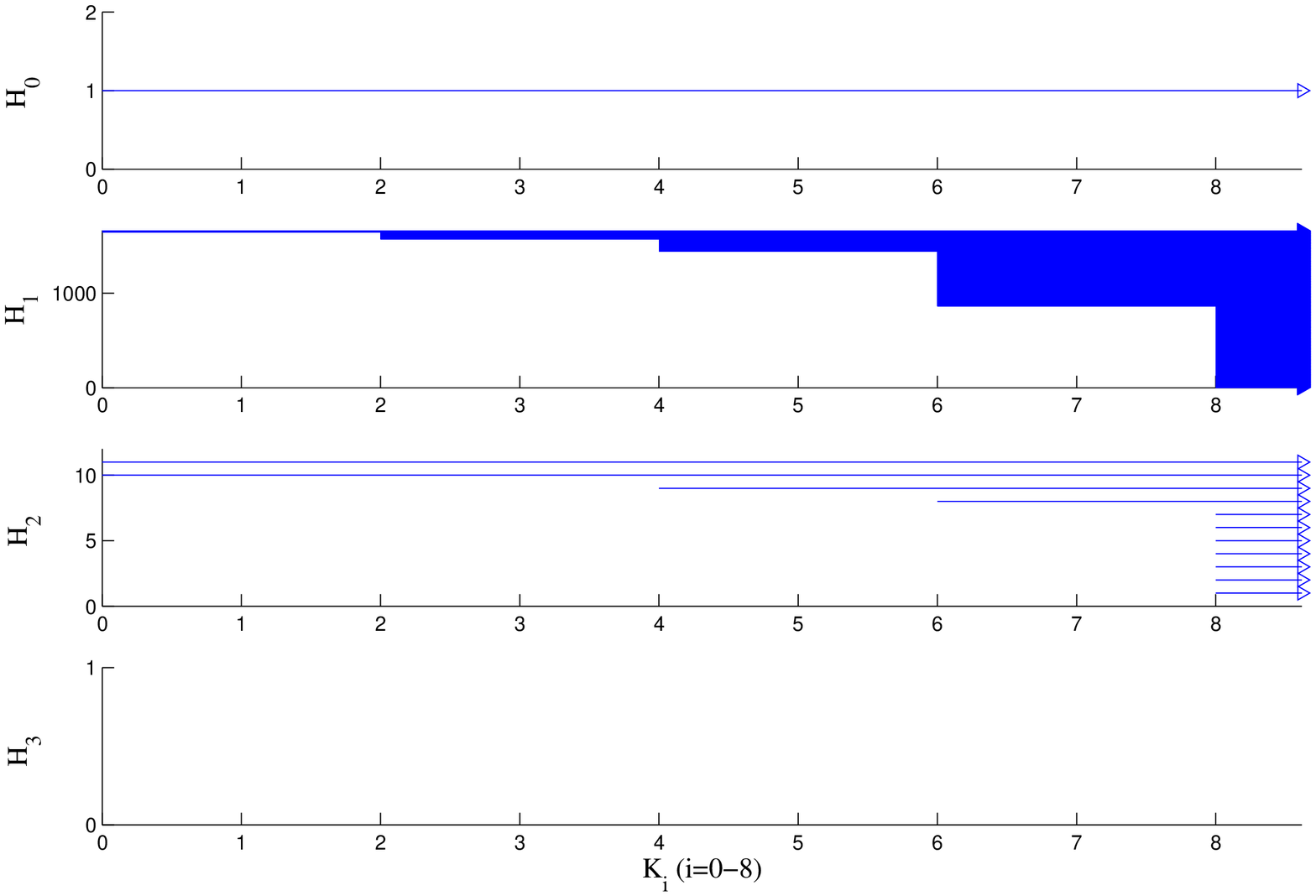}

Fig. 12 Barcode of the modular non-clustered network. The persistent
homology group with the highest rank is $H_{2}$.
\end{center}


\begin{thebibliography}{99}
\bibitem{barcelo} Barcelo H and Laubenbacher R Perspectives on A-homotopy
theory and its applications
\end{thebibliography}
\end{document}